\documentclass[twocolumn,showpacs,preprintnumbers,amsmath,amssymb]{revtex4}

\usepackage{graphicx}
\usepackage{dcolumn}
\usepackage{bm}

\begin{document}

\title{Fidelity optimization for holonomic quantum gates in dissipative 
environments}

\author{Daniele Parodi,$^{1,2}$ Maura Sassetti,$^{1,3}$
Paolo Solinas,$^{1,2}$ Paolo Zanardi $^{4}$, Nino Zangh\`{\i}$^{1,2}$}

\affiliation{
$^1$ Dipartimento di Fisica, Universit\`a di Genova, Via Dodecaneso 33, 16146 Genova, Italy\\
$^2$ Istituto Nazionale di Fisica Nucleare (Sezione di Genova), Via Dodecaneso 33, 16146 Genova, Italy \\
$^3$ INFM-CNR Lamia , Via Dodecaneso 33, 16146 Genova, Italy \\\\
$^4$ Institute for Scientific Interchange, Viale Settimio Severo 65, 
10133 Torino, Italy
}

\date{\today}

\begin{abstract}

 We  analyze the performance of holonomic quantum gates in semiconductor 
quantum dots, driven by ultrafast lasers, under the effect of a dissipative 
environment. 
The environment  is modeled as a thermal bath of  oscillators  linearly coupled
 with the electron states of the quantum dot.  Standard techniques 
make the problem amenable to a numerical treatment and allow one to determine 
the {\em fidelity} 
as a function of all the relevant physical parameters. As a consequence of our
 analysis, we  show that the disturbance of the environment can be 
(approximately) suppressed and   the performance of the gate  
optimized|provided that  the thermal bath is purely superhomic.   
We conclude by showing that such an optimization is impossible for ohmic 
environments.

\end{abstract}


\maketitle

\section{Introduction}

In the last years  holonomic quantum computation (HQC) has proved to be a 
viable and promising approach to quantum information processing  and quantum 
computation~\cite{HQC}. According to this approach,
quantum information is encoded in an $n$-fold degenerate 
eigenspace of a family of quantum Hamiltonians depending on dynamically 
controllable parameters. Recently,  concrete proposals for quantum computation 
have been put forward, for both Abelian  \cite{abelian} and
non-Abelian holonomies \cite{HQC_proposal,paper1-2}. 
The so-constructed  gates depends only  on global 
geometrical feature, e.g., the solid angle spanned in the parameter space, 
and because of this, it is believed that they are robust against errors  
affecting the physical parameters controlling the  gates themselves 
(e.g., laser pulses).
This expectation has been confirmed by recent investigations~
\cite{par_noise}.

An important open  problem is whether holonomic quantum gates are stable 
under the effect of the environment \cite{fuentes-guridi+thunstrom}.   
In this paper we argue that 
holonomic gates  have indeed a good performance when the effect of the 
environment is taken into account.  
We show this on the basis of a simple and idealized model which covers 
situations describing electron states in quantum dots ranging 
from excitons \cite{paper1-2} to optically active spin-degenerate ones 
\cite{troiani-molinari} (for a related investigation about environmental
effects on semiconductor based quantum gates see Ref. \cite{roszak}).
For such gates, the main source of dissipation is due to phonons with
{\it superohmic} spectral density \cite{weiss}.
It turns out that these gates manifest a rich structure when the 
``control parameters'' are changed.  
By  varying in a suitable way the adiabatic time,
the {\it superohmic} effect can be minimized and suppressed. 
For completeness,  we have extended our analysis to an ohmic environment.

In Sec. \ref{sec:HQC+env}, we review the HQC model,
introduce a model for the dissipative environment and write the master equation
to solve.
In Sec. \ref{sec:simulations}, we descibe the computer simulations and 
specify the kind of environment we consider. From the numerical results we 
deduce non-trivial behavior of the {\it fidelity} (the gate performance 
extimator) which allows us to 
minimize the decoherence effects of a {\it superohmic} environment.
We show that analogous minimization cannot be done with other kinds of 
environment ({\it ohmic}).

\section{HQC in a dissipative environment}
\label{sec:HQC+env}

The physical system we consider is constituted by three degenerate (or
quasi-degenerate) states ($|+\rangle$, $|-\rangle$, and $|0\rangle$)
optically connected to another state $|G\rangle$.
Every degenerate state is separately addressed by polarization or frequency 
selection with a laser. This model describes various quantum systems 
interacting with a laser radiation field; here, we deal with quantum states in 
semiconductor quantum dots such as excitons \cite{paper1-2} and 
spin-degenerate electron states \cite{troiani-molinari}.

The (approximate) Hamiltonian  modeling the  effect of the laser on the 
system is (for simplicity, $\hbar=1$) \cite{paper1-2}
\begin{equation}
\label{eq:system_hamiltonian}
      H_0(t)= \sum_{j=+,-,0} (\epsilon |j\rangle \langle j| +
      e^{-i \epsilon t} \Omega_j(t) |j\rangle \langle G|) + h.c.\; \mathbf{,}
\end{equation}
where $\Omega_j(t)$ are the time dependent Rabi frequencies depending on 
controllable parameters, such as the phase and intensity of the lasers, 
and $\epsilon$ is the energy of the degenerate electron states.
The Rabi frequencies can be modulated within the adiabatic time $t_{ad}$ 
(which coincides with the gating time) 
to produce a loop in the parameter space  and thereby realize the periodic 
condition $H_0(t_{ad})=H_0(0)$.

It should be observed that different loops in the parameter space produce 
different holonomic operators. Here
we consider two different sets of Rabi frequencies, that is, 
 two different choices of the time dependent functions $\Omega_j(t)$ 
in Eq. (\ref{eq:system_hamiltonian}).
According to the holonomic approach,
the corresponding unitary evolutions occurring in the  adiabatic time define 
the unitary transformations associated with the holonomic quantum gates  
\cite{paper1-2}.   
We shall hereafter refer to the gates defined by these two different sets of 
Rabi frequencies (which form a complete set of single qubit gates)
 as ``gate $1$'' and ``gate $2$''; they correspond, respectively, 
to the unitary operators $U_1 = e^{i \frac{\pi}{4} |+\rangle \langle +|}$ 
and $U_2 = e^{i \frac{\pi}{2} \sigma_y}$ 
(where $\sigma_y = i( |+\rangle \langle -| - |-\rangle \langle +|)$).

\begin{figure}[t]
  \begin{center}
    \includegraphics[height=5cm]{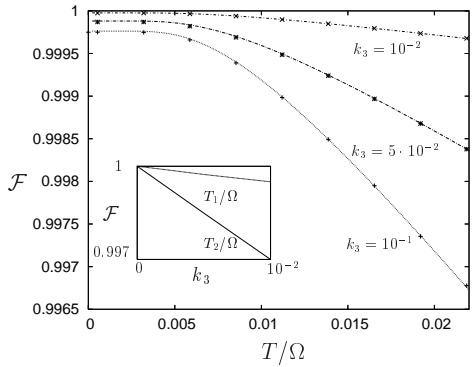}
    \caption{\label{fig:single_qubit} Fidelity $\mathcal{F}$
      for gate $1$ as a function of temperature for different $k_3$ 
      (expressed in  (meV)$^{-2}$)
      in the presence of a {\it superohmic} environment.
      The points are the computer simulation results and the curves have 
      the form $\mathcal{F}=1-\sum_{j=\pm} \eta_j \Gamma^j t_{ad}$
      ($\eta_-=3 \cdot 10^{-2}$ and $\eta_+= 0.7$).
      Inset: Fidelity $\mathcal{F}$ as a function of $k_3$ for two different 
      temperatures ($T_1/\Omega=1.6 \cdot 10^{-2}$ and 
      $T_2/\Omega=5 \cdot 10^{-3}$). Parameters:
      $\epsilon=1$~eV, $\Omega=25$~meV, 
      $t_{ad}=7.5$~ps 
      and $\omega_c=0.5$~meV.}
  \end{center}
\end{figure}

The Hamiltonian (\ref{eq:system_hamiltonian}) has four eigenstates:
two eigenstates $|B_{\pm}\rangle$ with time-dependent eigenvalues 
$\epsilon_\pm$ (called {\it bright states}) 
and two eigenstates  $|D_{1,2}\rangle$ with constant and degenerate eigenvalue 
$\epsilon$ (called {\it dark states}).
To construct a complete set of holonomic quantum gates it is sufficient 
to restrict Rabi frequencies $\Omega_j (t)$ such that the norm $\Omega$ of the 
vectors $\{\Omega_j (t)\}$  is time independent.    Under this condition,
it can be easily shown that the two dark states have 
energy $\epsilon$ and the two bright states have time independent energies 
%$\epsilon_\pm = (\epsilon \pm \sqrt{\epsilon^2 + 4 \Omega^2})/2$.
\begin{equation}
  \epsilon_\pm = (\epsilon \pm \sqrt{\epsilon^2 + 4 \Omega^2})/2
  \label{eq:energy_levels} .
\end{equation}
The adiabatic condition is simply $\Omega t_{ad} \gg 1$.

The environment is described as an ensemble of harmonic oscillators
linearly coupled to the system \cite{caldeira-leggett}, with total Hamiltonian
\begin{equation}
  \label{eq:bagno}
  H =        H_0(t) +
  \sum_{\alpha=1}^N (\frac{p^2_{\alpha}}{2 m_{\alpha}} + \frac{1}{2} 
m_{\alpha} \omega_{\alpha}^2 x_{\alpha}^2 + c_{\alpha} x_{\alpha} A)
\end{equation}
The interaction should break the degeneracy of the degenerate states, a 
condition that is easily fulfilled  by assuming   the operator $A$  to be of 
the form 
$A=\mbox{diag}(0,1,0,-1)$
in the basis $|G\rangle$, $|+\rangle$, $|0\rangle$, and $|-\rangle$.
(Eq.  (\ref{eq:bagno}) includes implicitly the standard renormalization 
term \cite{weiss}).

We now consider the time evolution of the reduced density matrix 
of the system, determined by the Hamiltonian (\ref{eq:bagno}). We rely on the 
standard methods of the ``master equation approach,''  according to which
the effect of the environment  is considered in the Born approximation,
and the environment is assumed to be at each time in its own thermal 
equilibrium state at temperature $T$.  
One has \cite{weiss}
{\small
\begin{eqnarray}
  \dot{\tilde{\rho}}(t) = - \int_0^t d\tau 
    \{ g(\tau) [ 
  \tilde{A} \tilde{A}^\prime \tilde{\rho}(t-\tau) - 
  \tilde{A}^\prime \tilde{\rho}(t-\tau) \tilde{A} ] \nonumber \\
    + g(-\tau) [
  \tilde{\rho}(t-\tau) \tilde{A}^\prime \tilde{A} - 
  \tilde{A} \tilde{\rho}(t-\tau) \tilde{A}^\prime ] \}
 \label{eq:non_markov_m_eq}
\end{eqnarray}
}
\hspace{-.2cm}where, $\tilde{\rho}(t)$ denotes the time evolution of the 
reduced density matrix of the system in  the interaction picture, e.g.,
$\tilde{\rho}(t) = U^{\dagger}(t,0) \rho U(t,0)$, where
$U(t,0)= \mbox{T} (\exp{-i \int_0^t d t^\prime H_0(t^\prime)}) $
and T is the time ordered product.
In the above equation $\tilde{A}$ and $\tilde{A}^\prime$ stand for 
$\tilde{A}(t)$ and $\tilde{A}(t-\tau)$ (again, the superscript tilde means 
time evolution in the interaction picture).
\begin{figure}[t]
  \begin{center}
   \includegraphics[height=5cm]{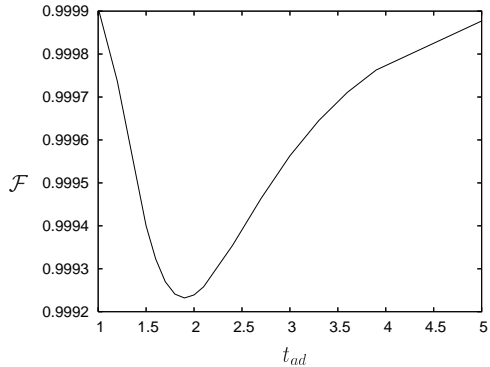}
    \caption{\label{fig:fid_ad} 
      Fidelity $\mathcal{F}$ for gate $1$ 
      subject to {\it superohmic} spectral density 
      as a function of $t_{ad}$ ($T/\Omega=8.5 \cdot 10^{-3}$, $k_3=10^{-1}$
      (meV)$^{-2}$ and $\Omega t_{ad}=280$). 
      The adiabatic time is normalized to $5$~ps. Parameters as in 
      Figure \ref{fig:single_qubit} except $t_{ad}$.
    }
  \end{center}
\end{figure}
In Eq. (\ref{eq:non_markov_m_eq}) the effect of the environment is 
included in the function (for simplicity, $k_B=1$)
\begin{equation}
  g(\tau) = \int^\infty_0 J(\omega) [ \coth(\frac{\omega}{2 T}) 
    \cos(\omega \tau) - i \sin(\omega \tau)] d\omega
  \label{eq:autocorrelation}
\end{equation}
where the spectral density $J(\omega)$ is defined in a standard way in terms 
of oscillator parameters in Eq. (\ref{eq:bagno}) and
its typical behavior, for 
physical environments in the low frequency regimes, is proportional to 
$\omega^s$, with $s\geq 1$ \cite{weiss}; 
the asymptotic decay  of the real part of $g(\tau)$
defines the characteristic memory time $\tau_E$ of the environment.

We solve Eq. (\ref{eq:non_markov_m_eq}) for an environment with 
memory time $\tau_E$ faster than the time scale of the variation of 
the density matrix $\tau_D$ 
(estimated self-consistently), so that
$\tilde{\rho}(t-\tau) \approx \tilde{\rho}(t)$ 
(Markov approximation). In this approximation, Eq.
(\ref{eq:non_markov_m_eq}) simplifies and  assumes a 
convenient form|for the numerical analysis we wish to perform|in the 
Schroedinger representation:
\begin{eqnarray}
  \dot{\rho}(t) = - i [H_0(t), \rho(t)] - \mathcal{L}(\rho)
  \label{eq:master_eq}
\end{eqnarray}
with 
\begin{eqnarray}
 \mathcal{L}(\rho) = \int^\infty_0 d\tau \{ 
 g(\tau)[ AA^{\prime \prime} \rho(t) - A^{\prime \prime}\rho(t) A] \nonumber \\ 
 + g(-\tau) [ \rho(t) A^{\prime \prime}A - A \rho(t)A^{\prime \prime}] \}   \,,
\label{eq:noise_term}
\end{eqnarray}
where $A^{\prime \prime} = U^{\dagger}(t-\tau,t)A U(t-\tau,t)$,
and, in the adiabatic approximation, $U(t-\tau,t) \approx \exp(i \tau H_0(t))$.
As usual, the upper extreme of integration is extended to infinity 
because the evolution time is much longer than $\tau_E$.
\begin{figure}[t]
  \begin{center}
   \includegraphics[height=5cm]{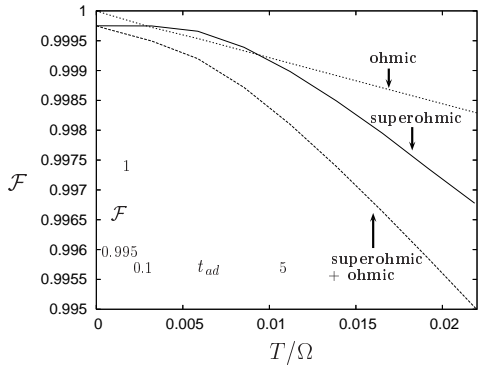}
    \caption{\label{fig:ohmic} 
      Fidelity $\mathcal{F}$ for gate $1$ 
      subject to {\it superohmic} (dashed line), {\it ohmic} (dot line) and 
      both (solid line) spectral densities as a function of $T/ \Omega$. 
      Inset: Fidelity $\mathcal{F}$ for {\it superohmic} plus
      {\it ohmic} environment as a function of $t_{ad}$ with 
      $\Omega t_{ad}=280$
      ($k_1= 4\cdot 10^{-4}$ and $k_3=10^{-1}$(meV)$^{-2}$). Parameters as in 
      Figure \ref{fig:single_qubit}.
}
  \end{center}
\end{figure}

\section{Computer simulations and results}
\label{sec:simulations}

In order to estimate how the environment affects the performance of the 
{\em ideal} holonomic gates determined by the ``dissipation-free'' 
Hamiltonian $ H_0(t)$, we use the standard notion of 
{\em fidelity} 
\begin{equation}
  \mathcal{F}= \sqrt{\langle \psi_{id}(t_{ad})| \rho(t_{ad}) 
    |\psi_{id}(t_{ad})\rangle}
  \label{eq:fidelity}
\end{equation}
where $|\psi_{id}(t_{ad})\rangle$ is the  state in which the initial (pure) 
state evolves, in the adiabatic time
$t_{ad}$, under the action only of  $ H_0(t)$,  while $\rho(t_{ad})$ is the 
solution of Eq. (\ref{eq:master_eq}),
computed at the same time, and for the {\em same} initial (pure) state.   
In order to avoid dependence on the latter we have taken a suitable average
on the initial states.
We make a sampling of the initial logical states (combination of the logical
states $|+\rangle$ and $|-\rangle$) on the Bloch sphere.
We add the possibility of an error in the preparation of the initial state 
with the population of the {\it non-logical} state $|0\rangle$.
This can be due to the imprecise control of the lasers.
The initial state has the form $\alpha |+\rangle + \beta |-\rangle +
\eta |0\rangle $ with $|\alpha|^2+|\beta|^2+|\eta|^2=1$ and
$|\eta|^2 = 0.1$. In the following, with a slight 
abuse of notation, we shall denote by the same symbol $  \mathcal{F}$  this 
averaged fidelity.

An essential ingredient of our analysis is the spectral density entering in Eq.
(\ref{eq:autocorrelation}).
For the electronic states in quantum dots the decoherence effects are 
principally due to phonons. Single phonon processes are described by 
{\it superohmic} spectral 
densities with $J(\omega) = k_3 \omega^3 e^{-(\omega/\omega_c)^2}$ 
\cite{alicki}. 
The high frequency cut-off $\omega_c$ is due to the planar confinement in 
the quantum dot.
The adimensional coupling constant $k_3$ allows the description
of different kinds phonon-carrier interactions in semiconductor materials
including deformation potential, piezoelectric and spin-orbit \cite{Nazarov}.
\begin{figure}[t]
  \begin{center}
   \includegraphics[height=5cm]{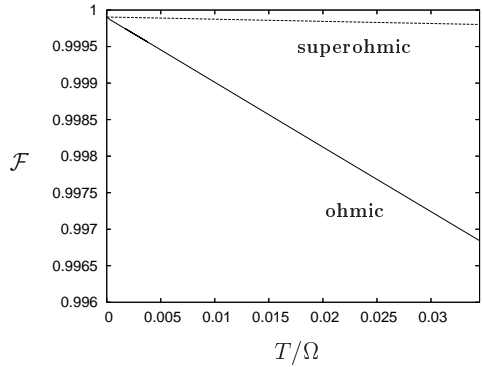}
    \caption{\label{fig:2qubs} 
      Fidelity for two qubit gate subject to {\it superohmic} and 
      {\it ohmic} spectral density as a function of 
      $T/ \Omega$ for $k_1=10^{-4}$ and $k_3=10^{-2}$(meV)$^{-2}$.  
      Parameters as in Figure \ref{fig:single_qubit} except $t_{ad}=0.8$~ns 
      and 
      $\Omega=0.2$~meV.
    }
  \end{center}
\end{figure}

The results of numerical simulations for the fidelity  $\mathcal{F}$ of gate 
$1$, as a functions of the temperature are shown in Fig. 
\ref{fig:single_qubit}:
at low temperature the fidelity is constant  and decreases linearly as the 
temperature increases.
To understand how this behavior comes about it is convenient  to express
the dissipative  {part} of $\mathcal{L}(\rho)$ in Eq. (\ref{eq:master_eq}) 
(in the dark-bright state basis) as  \cite{weiss}:
$$\sum_{kl}(\delta_{lm} \sum_r \Gamma^+_{nrrk}+
\delta_{nk} \sum_r \Gamma^-_{lrrk}
- \Gamma^-_{lmnk} - \Gamma^+_{lmnk}) \rho_{kl}\,.$$
Here, $\Gamma^{\pm}_{lmnk}=\Gamma^\pm_{nk} K_{lmnk}$ ,
with $K_{lmnk}$ depending only on laser parameters and
\begin{equation}
  \Gamma^\pm_{nk} (\omega_{nk})= J(\omega_{nk})(\coth (\frac{\omega_{nk}}{2 T})
 \mp 1 ) 
  \label{eq:rates}
\end{equation}
are the transition rates between $n$ and $k$ 
states due to phonons. In passing, we note that these are indeed the 
rates that  could be guessed 
by a straightforward application of Fermi's golden rule  to the
interaction terms of Eq.( \ref{eq:bagno}). Finally, observe that
the frequencies $\omega_{nk} = \epsilon_n - \epsilon_k$ represent the energy 
differences in the dark-bright space, i.e., $\epsilon_n=\epsilon, 
\epsilon_{\pm}$, where  $\epsilon_{\pm}$ are given by Eq. 
(\ref{eq:energy_levels} ); for $\epsilon \gg \Omega$, 
$\omega_{nk} = 0, \epsilon, \Omega^2/\epsilon$.
%and for $\epsilon \gg \Omega$, 
%$\omega_{nk} = 0, \epsilon, \Omega^2/\epsilon$. 

Equation (\ref{eq:rates}) shows that, with {\it superohmic} spectral density,
the only relevant transition is the one with $\omega_{nk} =\Omega^2/\epsilon$
giving the transition rates $\Gamma^\pm(\Omega^2/\epsilon)$.
In fact, for $\omega_{nk} = \epsilon$ the gaussian cut-off with 
$\omega_c \ll \epsilon$ produces negligible rates, and for degenerate 
dark state ($\omega_{nk} =0$) the 
rates vanish. These considerations, together with the explicit form of the 
rates  given by Eq. (\ref{eq:rates}) provide a compelling explanation of
the temperature behavior of the fidelity in Fig.\ref{fig:single_qubit}.   
Moreover, we have found that the numerical results are fitted by means of
the function
\begin{equation}
 \mathcal{F} = 1- t_{ad} \sum_{j=\pm} \eta_j \Gamma^j 
 \label{eq:universal_fidelity}
\end{equation}
(where $\eta_j$ are two real parameters). 
This behavior is also manifest  by considering the fidelity 
as a function of the coupling parameter $k_3$ (see inset in Fig. 
\ref{fig:single_qubit}). 
A similar dependence of the fidelity on the transition rate
has been found in Ref. \cite{roszak}.
Note that in that case the authors have only a single transition process
(absorption to higher states or emission to lower states)
while we need to take into account both 
absorption and emission processes for the transition to the higher state. 
This is due to the different master equation 
solved. In fact, they solve a strictly second 
order master equation, while our eq. (\ref{eq:master_eq}) is 
``self-consistent''. 
The similarity in the results is derived from the small value of the 
coupling constant $k_3$ but for higher values the two approximations 
diverge from each
other and the numerical results cannot be fitted by such an 
elementary function.

Eq. (\ref{eq:universal_fidelity}) is particularly important in that it allows 
one to predict how the fidelity behaves when the  parameters of the system are 
modified. 
To this end, first of all note that, by keeping the adiabatic parameter 
constant ($\Omega t_{ad} = \alpha =$ const), 
the rates $\Gamma^\pm$ become a non-trivial function of the adiabatic time. 
Then, by writing  the rates explicitly,  one obtains $\Gamma^\pm \propto 
1/t_{ad}^6 \coth (\alpha^2/(\epsilon t^2_{ad} T))
\exp(-(\alpha^2/(t^2_{ad} \epsilon \omega_c))^2)$. Thus, it follows from 
Eq.  (\ref{eq:fidelity})  that 
the fidelity should have a pronounced minimum as a function of $t_{ad}$.  
This behavior is confirmed by the computer simulations presented in Fig. 
\ref{fig:fid_ad}.
By varying $\Omega$ and $t_{ad}$ (e.g., by acting on the lasers) the position 
of the fidelity minimum in Fig. \ref{fig:fid_ad} can be shifted and
the effect of {\it superohmic} environment can be suppressed. 
It seems to us that this is an interesting result.

Before discussing the limitations of this result, we would like to comment on 
the approximations upon which our analysis 
relies. As we have already anticipated, in Eq. (\ref{eq:master_eq})
the Markov approximation is appropriate when the memory time 
$\tau_E$ is small with respect to the time scale of variation of the density 
matrix $\tau_D$.
Eq. (\ref{eq:autocorrelation}) leads to the estimate
$\tau_E \approx 1/(2 \pi T)$; while Eq. (\ref{eq:rates}), for a
{\it superohmic} spectral density and $T \ll \Omega^2 / \epsilon$, leads to 
$\tau_D = (\epsilon/\Omega^2)^3/k_3$.
Note that the conditions of validity of the  Markov approximation, 
$\tau_E < \tau_D$, readily translate into a   temperature regime, namely 
$T > T_M=k_3(\Omega^2/\epsilon)^3$.
With our choice of parameters, we have 
$1.2\cdot 10^{-4} \leq T_M/\Omega \leq 1.2\cdot 10^{-3}$
(depending on $k_3$ value), which, in our simulations, 
is a very low temperature.

Since the possibility of suppressing the {\it superohmic} 
effects is indeed surprising, one may wonder whether a similar possibility 
arises for more general environments, e.g. for ohmic environments. 
Though ohmic environments are typical of baths of conduction electrons 
\cite{weiss}, even for phonon baths, which are typically {\it superohmic}, 
it seems  possible that the  spectral density contains an ohmic part; this is 
presumably due to higher order contributions such as two phonon processes 
\cite{two_phonon}.

Be that as it may,  we found it interesting to extend our analysis to an
environment  with  the spectral density
$J(\omega) = k_1 \omega e^{-(\omega/\omega_c)^2}$ with $k_1 \ll 1$.
As is easily seen,  Eq. (\ref{eq:rates}) for 
ohmic rates leads to completely different results. This is due to
the presence, in the ohmic case, of transitions between degenerate states
which are absent in the case of a superhomic environment.  This difference has 
striking consequences: 
the  transitions between degenerate states give contribution to the rates 
(\ref{eq:rates}) which are linear in $T$  (while the transitions between 
non-degenerate states have the same temperature behavior as the 
{\it superohmic} case). 
This difference is confirmed by the computer simulations in Fig. 
\ref{fig:ohmic}, which shows the fidelity for the superohmic and 
ohmic environments, and the sum of the two contributions
as a function of $T$. This curve clearly shows that the presence of an 
ohmic environment changes dramatically the fidelity behavior, whence
it follows the impossibility of  extending to the ohmic case
the results previously obtained changing the adiabatic time.
This conclusion becomes very clear if one compares Fig. \ref{fig:fid_ad} with 
the inset in Fig. \ref{fig:ohmic}. It is not possible anymore to 
optimize the fidelity by changing the 
parameters. 

Before concluding, we briefly mention three points (see \cite{long_version} 
for a thorough discussion).
First, the computer simulation for gate $2$ confirms the results found for 
gate $1$. Second, our analysis extends (almost straightforwardly) to the 
two-qubit gate proposed in Ref. \cite{paper1-2}: Fig. \ref{fig:2qubs} shows 
the preliminary results for the two-qubit gate (for a non random initial state)
 as a function of temperature and coupling 
constant for  {\it superohmic} and {\it ohmic} environments;
the behavior of the fidelity is analogous to that of the one qubit case 
(Fig.~\ref{fig:ohmic}). Third, for the ohmic case, a careful study of the 
(relatively) high temperature behavior  shows the limitations of the 
Lindblad approximation for the reduced dynamics.

\section{Conclusions}
\label{sec:conclusions}

To sum up, the upshot of our analysis is twofold.
The good news is that it is possible to optimize the fidelity for 
the kind of environment which is usually considered for electron states 
in quantum dots|a 
superohmic environment caused by electron-phonon interactions.
The bad news is  that such optimization does not go through an
ohmic environment, e.g., produced by the {\em same} superohmic phonon bath 
through two phonon processes. 
Thus, particular attention should be paid in modeling the 
environment,  since the presence of a weak  ohmic environment 
dramatically changes the holonomic gate performance.
For these reasons
it is crucial to  dispose of  experimental investigations on the nature 
of the environmental spectral densities in semiconductor quantum dots.


\begin{thebibliography}{99}

\bibitem{HQC}
  P. Zanardi, and M. Rasetti, Phys. Lett. A {\bf 264}, 94 (1999);

\bibitem{abelian}
  J. A. Jones, V. Vedral, A. Ekert, and G. Castagnoli, 
  Nature (London) {\bf 403}, 869 (2000);
  G. Falci  {\it et al.}, Nature (London) {\bf 407}, 355 (2000).

\bibitem{HQC_proposal}
   R.G. Unanyan, B.W. Shore, and  K. Bergmann,
   Phys. Rev. A {\bf 59}, 2910 (1999);
   L.-M. Duan, J.I.  Cirac, and P. Zoller, Science {\bf 292}, 1695 (2001);
   L. Faoro, J. Siewert, and R. Fazio, Phys. Rev. Lett. {\bf 90}, 
   028301 (2003);
   I. Fuentes-Guridi {\it et al.}, Phys. Rev. A {\bf 66}, 022102 (2002);
   A. Recati {\it et al.}, Phys. Rev. A {\bf 66}, 032309 (2002).

\bibitem{paper1-2}
  P. Solinas, P. Zanardi, N. Zangh\`{\i}, and F. Rossi,
  Phys. Rev. B {\bf 67}, 121307 (2003)(R);
  P. Solinas, P. Zanardi, N. Zangh\`{\i}, and F. Rossi, 
  Phys. Rev. A {\bf  67}, 062315 (2003).

\bibitem{par_noise}
 A. Carollo, I. Fuentes-Guridi, M. F. Santos, and V. Vedral, 
 Phys. Rev. Lett. {\bf 90}, 160402 (2003);
 G. De Chiara and G.M. Palma, Phys. Rev. Lett. {\bf 91}, 090404 (2003);
 A. Carollo, I. Fuentes-Guridi, M. F. Santos, and V. Vedral, 
 Phys. Rev. Lett. {\bf 92}, 020402 (2004);
 V.I. Kuvshinov and A.V. Kuzmin, Phys. Lett. A, {\bf 316}, 391 (2003);
 P. Solinas, P. Zanardi, and N. Zangh\`{\i}
 Phys. Rev. A 70, 042316 (2004);
 %L.-X. Cen, and Paolo Zanardi Phys. Rev. A {\bf 70}, 052323 (2004);  
 S.-L. Zhu and P.Zanardi, Phys. Rev. A {\bf 72}, 020301 (2005)(R).
 
\bibitem{fuentes-guridi+thunstrom}
  A. Nazir, T.P. Spiller, and W. J. Munro, Phys. Rev. A {\bf 65}, 
  042303 (2002); 
  A. Blais, and A.-M. S. Tremblay, Phys. Rev. A {\bf 67}, 012308 (2003);
  I. Fuentes-Guridi, F. Girelli and E. R. Livine, Phys. Rev. Lett. {\bf 94}, 
  020503 (2005); 

\bibitem{troiani-molinari}
F. Troiani, E. Molinari and U. Hohenester, Phys. Rev. Lett. {\bf 90}, 206802 
(2003).

\bibitem{roszak} K. Roszak {\it et al.}
Phys. Rev. B {\bf 71}, 195333 (2005).

\bibitem{weiss}
U. Weiss, {\it Quantum dissipative systems}, World Scientific, Singapore 1999.


\bibitem{caldeira-leggett}
 A. O. Caldeira and A. J. Leggett, Phys. Rev. Lett. {\bf 46}, 211 (1981).

\bibitem{alicki}
  R. Alicki  {\it et al.},  Phys. Rev. A {\bf 70}, 010501(R) (2004).

\bibitem{Nazarov}
  A. V. Khaetskii and Y. V. Nazarov, Phys. Rev. B 61, 12642  (2000)

\bibitem{two_phonon}
 F. Napoli, M. Sassetti, and U. Weiss, Physica B {\bf 202}, 80 (1994).

\bibitem{long_version}
  D. Parodi {\it et al.} in preparation.

\end{thebibliography}
\end{document}